\documentclass[aps, prd, twocolumn, superscriptaddress, nofootinbib, floatfix, noshowpacs]{revtex4-2}
\pdfoutput=1
\newif\ifcomment
\usepackage{amsmath,amssymb}
\usepackage{color,url}
\usepackage{listings}
\usepackage{slashed}
\usepackage[pdftex]{graphicx}
\usepackage{epstopdf}
\usepackage{epsfig}
\usepackage{grffile}
\usepackage{relsize}
\usepackage{float}
\graphicspath{{./img/}}
\usepackage{soul}

\usepackage{CJKutf8}

\usepackage[mathscr,scaled=1.15]{urwchancal}
\DeclareFontFamily{OT1}{pzc}{}
\DeclareFontShape{OT1}{pzc}{m}{it}%
{<-> s * [1.15] pzcmi7t}{}
\DeclareMathAlphabet{\mathpzc}{OT1}{pzc}{m}{it}

\definecolor{purple}{rgb}{0.5,0,0.5}
\definecolor{blue}{rgb}{0.0,0,0.9}
\definecolor{prdblue}{rgb}{0.133,0.118,0.498}
\usepackage[colorlinks=true, pdfstartview=FitV, linkcolor=prdblue, citecolor= prdblue, urlcolor=prdblue]{hyperref}

\RequirePackage{comment}

\begin{document}
\begin{CJK*}{UTF8}{gbsn}

\title{$\,$\\[-6ex]\hspace*{\fill}{\normalsize{\sf\emph{Preprint no}.\
NJU-INP 124-26}}\\[1ex]
Lepton Flavour Universality in Semileptonic $\bar B_{(s)} \to D_{(s)}^{(\ast)}$ Transitions}

\author{Yu-Chen Hu（胡宇琛)%
       $^{\href{https://orcid.org/0009-0004-8330-1454}{\textcolor[rgb]{0.00,1.00,0.00}{\sf ID}}}$}
\affiliation{School of Physics, \href{https://ror.org/01rxvg760}{Nanjing University}, Nanjing, Jiangsu 210093, China}
\affiliation{Institute for Nonperturbative Physics, \href{https://ror.org/01rxvg760}{Nanjing University}, Nanjing, Jiangsu 210093, China}

\author{Cong-Cong Ye (叶聪聪)%
       $^{\href{https://orcid.org/0009-0004-4621-1732}{\textcolor[rgb]{0.00,1.00,0.00}{\sf ID}}}$}
\affiliation{School of Physics, \href{https://ror.org/01rxvg760}{Nanjing University}, Nanjing, Jiangsu 210093, China}
\affiliation{Institute for Nonperturbative Physics, \href{https://ror.org/01rxvg760}{Nanjing University}, Nanjing, Jiangsu 210093, China}

\author{Craig D. Roberts%
       $^{\href{https://orcid.org/0000-0002-2937-1361}{\textcolor[rgb]{0.00,1.00,0.00}{\sf ID}}}$}
\email[]{cdroberts@nju.edu.cn}
\affiliation{School of Physics, \href{https://ror.org/01rxvg760}{Nanjing University}, Nanjing, Jiangsu 210093, China}
\affiliation{Institute for Nonperturbative Physics, \href{https://ror.org/01rxvg760}{Nanjing University}, Nanjing, Jiangsu 210093, China}

\author{Zhao-Qian Yao (姚照千)%
       $^{\href{https://orcid.org/0000-0002-9621-6994}{\textcolor[rgb]{0.00,1.00,0.00}{\sf ID}}}\,$}
\affiliation{\href{https://ror.org/01zy2cs03}{Helmholtz-Zentrum Dresden-Rossendorf}, Bautzner Landstra{\ss}e 400, D-01328 Dresden, Germany}
%

\date{
2026 October 12 
}
\begin{abstract}
Using continuum Schwinger function methods, we deliver parameter-free pointwise predictions for the complete set of twelve Standard Model (SM) $\bar B_{(s)} \to D_{(s)}^{(\ast)}$
semileptonic transition form factors on the entire kinematically accessible domain.  Therewith, we calculate SM theory values for the decay-rate ratios of $\tau$ and $\mu$ final states:
$\{ R(D) \, , \, R(D)^\ast \} = \{ 0.290(11) \, , \, 0.267(5) \}$, 
$\{ R(D_{s}) \, , \, R(D_{s}^{\ast}) \}  = \{ 0.283(11) \, , \, 0.251(5) \}$.
The former is consistent with an existing average of SM estimates, but disagrees with the experimental average by $2.8\sigma$.
Comparing transitions with light quark and strange quark spectators, we find a $2.3\sigma$ signal for violation of SU$(3)$ flavour symmetry, expressed at the 5\% level in $D_{(s)}^\ast$ final states.  It is induced by Higgs boson couplings into the strong interaction and modulated by phenomena deriving from emergent features of the strong interaction.
\end{abstract}

\maketitle

\end{CJK*}

\noindent\emph{1.\ Introduction}\,---\,%
In the Standard Model (SM) of particle physics the interaction between weak gauge bosons and each lepton flavour is determined by the same coupling.  This lepton flavour universality (LFU) is only broken by lepton mass terms generated by flavour-dependent Higgs boson couplings.  In an era of high precision experiments, LFU can be checked: any violation is a clear signal of beyond-SM physics.

\begin{figure*}[t]
\centerline{%
\includegraphics[clip, width=0.8\textwidth]{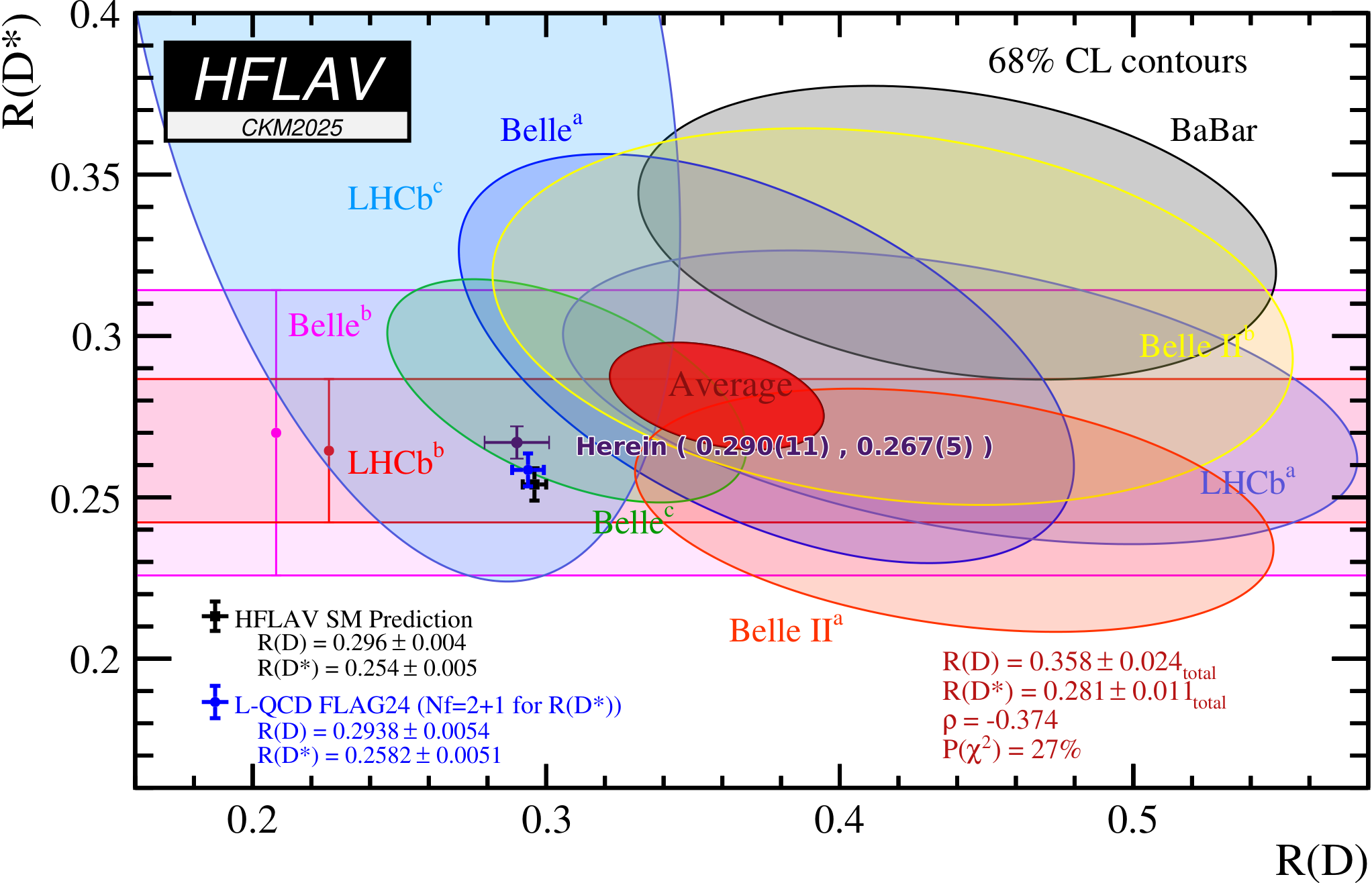}}
\vspace*{-2ex}

\caption{\label{F1F}
$\{R(D) , R(D^\ast)\}$ -- selected available measurement constraints and calculations as reported by the heavy flavor averaging group \cite[HFLAV]{HFLAV:2024ctg}.
The SM prediction (black cross) differs from the experimental average by $\approx 3.3\sigma$.  
A result based solely on results obtained using lattice-regularised QCD is also drawn (blue cross).
The purple point labelled ``Herein'' is our prediction, Eq.\,\eqref{RDDs}, to be described below.
(Image adapted from Ref.\,\cite[HFLAV]{HFLAV:2024ctg}:  ``herein'' point and associated text added.)
}
\end{figure*}

Especially useful transitions are those which are semileptonic and involve a single hadron with one heavy quark in both the initial and final state because, in such cases, some aspects of the theoretical analysis are simplified.  Pertinent examples are 
$\bar B_{(s)} \to D_{(s)}^{(\ast)+} \ell^- \bar\nu_\ell$, 
which involve the weak $b \to c$ transition, where $b$, $c$ are the bottom/beauty and charm quarks respectively.  They are of keen interest because measurements suggest some deviations from SM predictions that exceed three standard deviations \cite{Klaver:2024xdy}. 
Crucially, in the SM, the structure of such mesons is supposed to be described by quantum chromodynamics (QCD); hence, sound predictions of transition measurables require the use of techniques capable of reliably solving the hadron bound state problem.

In testing LFU, it is common to consider the following branching fraction ratios:
\begin{equation}
R(D_{(s)}^{(\ast)}) = \frac{{\mathpzc B}(\bar B_{(s)} \to D_{(s)}^{(\ast)} \tau \bar\nu_\tau)}{{\mathpzc B}(\bar B_{(s)} \to D_{(s)}^{(\ast)} \ell \bar\nu_\ell)} ,
\label{BFratios}
\end{equation}
where the $\tau$-lepton final state is the numerator and $\ell$ represents either an electron, $e^-$, or muon, $\mu^-$.  
This is because the Cabibbo-Kobayashi-Maskawa (CKM) matrix element, $|V_{cb}|$, which measures the strength of $b\to c$ weak transitions, cancels in this ratio, as do some theoretical uncertainties that affect the hadronic current.  Similarly, in measurements, a range of experimental uncertainties also cancel.  Altogether, these things make such ratios a (fairly) clean testbed for LFU.

A good amount of information is now available on the ratios $R(D)$, $R(D^\ast)$.
Figure\,\ref{F1F} displays a contemporary snapshot of the comparison between SM theory and experiment: evidently, there is a roughly $3.3\sigma$ discrepancy, which owes largely to $R(D)$.
For more than a decade \cite{BaBar:2012obs}, this has been interpreted as a hint of LFU-violation.

Nothing is known empirically about the ratios $R(D_s)$, $R(D_s^\ast)$: for the $D_{(s)}^\ast$ final states, only the branching fraction to $\mu \bar\nu_\mu$ is measured \cite{ParticleDataGroup:2024cfk}.
Independent, data-unbiased estimates are available from an array of theoretical tools \cite{Kang:2018jzg, Zhou:2019stx, Bordone:2019guc, Harrison:2023dzh, Ivanov:2026pul}.  
The unweighted mean of available estimates is
\begin{equation}
\{ R(D_{s}) \, , \, R(D_{s}^{\ast}) \} = \{ 0.295(10) \, , \, 0.251(09) \} \; .
\label{ResRDsDss}
\end{equation}

Of special interest here are the differences, if any, between 
$\{ R(D_{s}) \, , \, R(D_{s}^{\ast}) \} $
and
$\{ R(D) \, , \, R(D^\ast) \}$.
With $|V_{cb}|$ eliminated in the ratios, then such differences are an expression of SU$(3)$ light-quark flavour symmetry breaking in the strong interaction because the only obvious distinctions in the final states are ${\mathpzc l} = (u,d) \to s$, where $u$, $d$, $s$ denote up, down, strange quarks.  
If one refers only to the HFLAV SM prediction for $\{ R(D) \, , \, R(D^\ast) \}$ -- see Fig.\,\ref{F1F}, then it is practically identical to the unweighted-mean $s$-quark values in Eq.\,\eqref{ResRDsDss}.  Since the $D_s^{(\ast)}-D^{(\ast)}$ mass difference is $\approx 5$\% of the $D_s^{(\ast)},D^{(\ast)}$ mean masses, so marking a natural size for SU$(3)$ symmetry breaking, 
it would be surprising to find equality between $\{ R(D_{s}) \, , \, R(D_{s}^{\ast}) \} $ 
and
$\{ R(D) \, , \, R(D^\ast) \}$ in Nature.

SU$(3)$-flavour symmetry breaking measures interference between the effect of Higgs boson couplings into QCD and emergent hadron mass (EHM) in Nature.  EHM is a novel phenomenon, the possibility of which was identified long ago \cite{Schwinger:1962tn, Cornwall:1981zr}, yet whose character and corollaries are only now beginning to be elucidated via strong interaction experiment and theory \cite{Roberts:2021nhw, Ding:2022ows, Quintans:2022utc, Ferreira:2023fva, Mezrag:2023nkp, Achenbach:2025kfx, Lu:2025bnm, Binosi:2026tre}.

This study will use continuum Schwinger function methods (CSMs) to deliver SM predictions for the branching fractions described above.
It will exploit a contemporary understanding of EHM and simultaneously unify those results with predictions for, \emph{inter alia}, analogous ratios associated with semileptonic $B_c \to \eta_c, J/\psi$ transitions.
The latter are made topical by a measurement \cite{LHCb:2017vlu} that delivers a result for the associated LFU ratio, $R(J/\psi)$, that is $2\sigma$ away from SM theory. 

\smallskip

\noindent\emph{2.\ Transition Form Factor Definitions}\,---\,%
%
The following transition matrix elements are the focus of this study
($\mathsf P= \bar B_{(s)}$, $\mathsf P_f = D_{(s)}^+$, $\mathsf V_f = D_{(s)}^{(\ast)+}$):
{\allowdisplaybreaks\begin{subequations}
\label{EqMEs}
\begin{align}
M_\mu^{\mathsf P\to \mathsf P_f}& (P,Q) =\langle \mathsf P_f(p_{\mathsf P_f}) | \bar c i\gamma_\mu b |\mathsf P(k)\rangle \nonumber \\
& \hspace*{-1em}  = f_+^{\mathsf P_f}(t) \, T_{\mu\nu}^Q P_\nu  + f_0^{\mathsf P_f}(t)\, \tfrac{P\cdot Q}{Q^2} Q_\mu \,,\\
M_{\mu;\lambda}^{\mathsf P \to \mathsf V_f}&(P,Q) =\langle \mathsf V_f^\lambda(p_{\mathsf V_f};\lambda) | \bar c i(\gamma_\mu - \gamma_\mu \gamma_5) b |\mathsf P(k)\rangle \nonumber \\
& \hspace*{-1em} =  2 m_{\mathsf V_f} \tfrac{Q_\mu \epsilon^\lambda\cdot Q}{Q^2} A_0^{\mathsf V_f}(t)
+  [m_{\mathsf P}+m_{\mathsf V_f}]T_{\mu\nu}^Q \epsilon_\nu^\lambda \, A_1^{\mathsf V_f}(t)  \nonumber \\
&  + [P_\mu + Q_\mu \tfrac{m_{\mathsf P}^2-m_{\mathsf V_f}^2}{Q^2}]
\frac{\epsilon^\lambda\cdot Q \, A_2^{\mathsf V_f}(t)}{m_{\mathsf P}+m_{\mathsf V_f}} \nonumber \\
& 
+ \varepsilon_{\mu\nu\rho\sigma}\epsilon_\nu^\lambda k_\rho p_{\mathsf V_f}
\frac{2 V^{\mathsf V_f}(t)}{m_{\mathsf P}+m_{\mathsf V_f}}\,,
\end{align}
\end{subequations}
where $Q^2T_{\mu\nu}^Q = Q^2\delta_{\mu\nu} - Q_\mu Q_\nu$,
$P = (k+p_{\mathsf P_f, \mathsf V_f})$,
$Q = (p_{\mathsf P_f, \mathsf V_f}-k)$,
with $k^2 = -m_{\mathsf P}^2$, $p_{\mathsf P, \mathsf V_f}^2=-m_{\mathsf P, \mathsf V_f}^2$;
$\epsilon_\nu^\lambda(p_{\mathsf V_f})$ is a polarisation four-vector, with $\sum_{\lambda=1}^3\epsilon_\nu^\lambda(p_{\mathsf V_f})\epsilon_\mu^\lambda(p_{\mathsf V_f}) = T_{\mu\nu}^{p_{\mathsf V_f}}$;
the squared-momentum-transfer is $t=-Q^2$;
and
$t_\pm^M = (m_{\mathsf P} \pm m_{\mathsf M})^2$,
$\mathsf M=\mathsf P_f, \mathsf V_f$.  ($t_-$ is the largest accessible value of $t$ in the identified physical decay process.)
The absence of kinematic singularities in Eqs.\,\eqref{EqMEs} is guaranteed by symmetries, which entail:
\begin{subequations}\label{EQsymmetries}
\begin{align}
f_+^{\mathsf P_f}(0)& =f_0^{\mathsf P_f}(0)\,,\\
A_0^{\mathsf V_f}(0) & = 
\tfrac{m_{\mathsf P} + m_{\mathsf V_f}}{2m_{\mathsf V_f}} A_1^{\mathsf V_f}(0) - \tfrac{m_{\mathsf P} - m_{\mathsf V_f}}{2m_{\mathsf V_f}} A_2^{\mathsf V_f}(0)\,.
\end{align}
\end{subequations}
}

The scalar functions in Eqs.\,\eqref{EqMEs} are the semileptonic transition form factors.
They are nonperturbative quantities that express all effects of hadron structure on the transitions.  
No study of LFU is reliable unless it employs techniques that deliver robust information on these hadron structure factors.

\smallskip

Branching fractions are calculated from the $\mathsf P \to \mathsf M \mathpzc l \bar \nu_{\mathpzc l}$, $\mathpzc l=\tau, \ell$, differential decay widths:
\begin{align}
\label{dGdt}
\left.\frac{d\Gamma}{dt}\right|_{\mathsf P \to \mathsf H \mathpzc l \bar \nu_{\mathpzc l}} &
= \frac{G_F^2 |V_{cb}|^2}{192\pi^3 m_{\mathsf P}^3}
\lambda(m_{\mathsf P},m_{\mathsf M},t) \frac{(t-m_{\mathpzc l}^2)^2}{t^2}\, {\mathpzc H}^2,
\end{align}
where:
$\lambda(m_{\mathsf P},m_{\mathsf M},t)^2 = (t_+ - t)(t_- - t)$;
and 
$G_F = 1.166 \times 10^{-5}\,$GeV$^{-2}$,
$|V_{cb}| = 0.0407(13)$ \mbox{\cite{ParticleDataGroup:2024cfk}};
{\allowdisplaybreaks
\begin{align}
{\mathpzc H}^2 & = (H_+^2 + H_-^2 + H_0^2)(1+\tfrac{m_{\mathpzc l}^2}{2t})
+ \tfrac{3 m_{\mathpzc l}^2}{2 t} H_t^2,
\end{align}
$m_{\mathpzc l}^2\leq t \leq t_-$, $m_{\mathpzc l}$ is the lepton mass.
For $\mathsf M=\mathsf P_f$, $H_\pm \equiv 0$,
\begin{equation}
H_0 = \lambda(m_{\mathsf P},m_{\mathsf P_f},t) f_+(t)\,,\;
H_t = (m_{\mathsf P}^2 - m_{\mathsf P_f}^2) f_0(t)\,;
\end{equation}
whereas when $\mathsf M= \mathsf V_f$,
\begin{subequations}
\label{Hpsi}
\begin{align}
\tfrac{1}{\surd t}H_\pm & = (m_{\mathsf P}+m_{\mathsf V_f})A_1 (t) \mp \frac{\lambda(m_{\mathsf P},m_{\mathsf V_f},t)}{m_{\mathsf P}+m_{\mathsf V_f}} V(t) \,,\\
H_0 & = \frac{1}{2 m_{\mathsf V_f}} \left[
(m_{\mathsf P}^2-m_{\mathsf V_f}^2 -t)(m_{\mathsf P}+m_{\mathsf V_f}) A_1(t) \right.   \nonumber \\
& \quad \left. - \frac{\lambda(m_{\mathsf P},m_{\mathsf V_f},t)^2}{m_{\mathsf P}+m_{\mathsf V_f}} A_2(t)\right],\\
H_t & =\lambda(m_{\mathsf P},m_{\mathsf V_f},t)\, A_0(t)\,.
\end{align}
\end{subequations}
After integrating Eq.\,\eqref{dGdt} to obtain the required partial widths, one quotes the branching fractions, ${\mathpzc B}_{\mathsf P\to \mathsf M{\mathpzc l} \nu_{\mathpzc l}}$, with respect to the total widths listed in Table~\ref{BranchingFracs}.
}


\begin{table}[t]
\caption{ \label{BranchingFracs}
Predicted semileptonic partial decay widths and branching fractions (herein), along with experiment comparisons (where available).
In calculating the widths and fractions, we use \cite[PDG]{ParticleDataGroup:2024cfk}:  $|V_{cb}| = 0.0407(13)$; 
$\tau_{\bar B^0} = 1.5113(0027)\,$ps, 
$\tau_{\bar B_{s}}= 1.515(006)\,$ps.
(These values have no effect on the branching fraction ratios defined in Eq.\,\eqref{BFratios}.%
)
}
\begin{center}
\begin{tabular*}
{\hsize}
{
l@{\extracolsep{0ptplus1fil}}
|c@{\extracolsep{0ptplus1fil}}
l@{\extracolsep{0ptplus1fil}}
l@{\extracolsep{0ptplus1fil}}
l@{\extracolsep{0ptplus1fil}}
l@{\extracolsep{0ptplus1fil}}}\hline\hline
transition$\rule{0em}{5ex}\ $
& lepton
& \shortstack{$\Gamma_{\rm herein}$\\$[10^{11}\!/{\rm s}]$}
& \shortstack{$\Gamma_{\rm PDG}$\\$[10^{11}\!/{\rm s}]$}
& \shortstack{$\mathcal{B}_{\rm herein}$\\$[\%]$}
& \shortstack{$\mathcal{B}_{\rm PDG}$\\$[\%]$}
\\\hline
$B\to D$
& $e$
& $0.142(11)$
& $0.140(4)$
& $2.14(17)$
& $2.12(06)$
\\
& $\mu$
& $0.141(11)$
& $0.140(4)$
& $2.13(17)$
& $2.12(06)$
\\
& $\tau$
& $0.0409(30)$
& $0.0476(99)$
& $0.618(45)$
& $0.72(15)$
\\ \hline
$B_s\to D_s$
& $e$
& $0.155(12)$
&
& $2.34(19)$
&
\\
& $\mu$
& $0.154(12)$
& $0.152(14)$
& $2.33(18)$
& $2.31(21)$
\\
& $\tau$
& $0.0436(32)$
&
& $0.661(48)$
&
\\ \hline
$B\to D^*$
& $e$
& $0.286(25)$
& $0.324(8)$
& $4.32(38)$
& $4.90(12)$
\\
& $\mu$
& $0.284(25)$
& $0.324(8)$
& $4.30(38)$
& $4.90(12)$
\\
& $\tau$
& $0.0760(62)$
& $0.0993(53)$
& $1.15(10)$
& $1.50(08)$
\\ \hline
$B_s\to D_s^*$
& $e$
& $0.343(29)$
& 
& $5.20(44)$
& 
\\
& $\mu$
& $0.342(29)$
& $0.343(33)$
& $5.18(44)$
& $5.2(5)$
\\
& $\tau$
& $0.0856(67)$
& 
& $1.30(11)$
& 
\\
\hline\hline
\end{tabular*}
\end{center}
\end{table}

\smallskip

\noindent\emph{3.\ Transition Matrix Elements}\,---\,%
We compute the transition form factors at leading-order (rainbow-ladder, RL) in the widely used and extensively benchmarked CSM truncation scheme introduced in Refs.\,\cite{Munczek:1994zz, Bender:1996bb}.
Today, this is understood to deliver reliable approximations for the Schwinger functions which appear in the matrix elements that describe measurable processes.   There are two reasons.
(\emph{i}) In QCD, symmetries are crucial; and RL truncation is built such that those symmetries germane to a given calculation are preserved.  This is discussed in Refs.\,\cite{Munczek:1994zz, Bender:1996bb, Bicudo:2001jq, Ding:2019lwe} and many citations thereof.
(\emph{ii}) A typical $n$-point Schwinger function describes $n$-parton/particle interactions. With increasing $n$, the likelihood of $n$ partons ``locating'' each other, in order to interact, diminishes rapidly. 
Furthermore, QCD's process-independent interaction strength, $\hat\alpha(r)$ \cite{Cui:2019dwv}, vanishes with decreasing inter-parton spacetime separation, $r$ -- this is asymptotic freedom \cite{Politzer:2005kc, Wilczek:2005az, Gross:2005kv}, and it is bounded above at long range, \emph{viz}.\ $\hat\alpha(r\gg 1/m_\pi) \lesssim \pi$.
Consequently, with increasing $n$, the contribution of higher $n$-point functions to any matrix element is rapidly suppressed.

{\allowdisplaybreaks 
In RL truncation (${\mathpzc q}_{v}={\mathpzc q}-v$):
\begin{align}
\nonumber
&M_\mu^{\mathsf P \to \mathsf M}(P,Q)  = 2N_c {\rm tr}\int\frac{d^4 {\mathpzc q}}{(2\pi)^4}
S_b({\mathpzc q}_{k}) \Gamma_{\mathsf P}({\mathpzc q}_{k/2};k) \\
& \qquad \times S_{{\mathpzc l},s}({\mathpzc q}) \bar\Gamma_{\mathsf M}({\mathpzc q}_{p/2};-p) S_c({\mathpzc q}-p)
i {\mathpzc W}_\mu^{cb}({\mathpzc q}_p,{\mathpzc q}_k) \,,
\label{dMDs}
\end{align}
where $N_c=3$ and the trace is over spinor indices.
Three types of matrix-valued function appear in Eq.\,\eqref{dMDs}.  
The propagators for the dressed-quarks involved in the transition process, \emph{viz}.\ $S_f({\mathpzc q})$, $f=b, c, s, {\mathpzc l}$; 
the Bethe-Salpeter amplitudes for the mesons involved: $\Gamma_{\mathsf P, \mathsf M}$; 
and the dressed $b\to c$ weak transition vertex: ${\mathpzc W}_\mu^{bc}$.  
Each of these quantities can be computed once the kernel of the RL Bethe-Salpeter equation is specified.
Importantly, Eq.\,\eqref{dMDs} guarantees the identities in Eq.\,\eqref{EQsymmetries}.
}

Extensive analyses of QCD gauge sector dynamics have revealed that a realistic RL kernel is expressed in the following parametrisation of the product of effective charge and gluon propagator \cite{Qin:2011dd, Qin:2011xq, Binosi:2014aea, Binosi:2016wcx, Yao:2024ixu}:
\begin{align}
\label{defcalG}
 {\mathpzc d}(y) & =
 \frac{2\pi}{\omega^4} D e^{-y/\omega^2} + \frac{\pi \gamma_m \mathcal{F}(y)}{\tfrac{1}{2}\ln\big[ \tau+(1+y/\Lambda_{\rm QCD}^2)^2 \big]}\,,
\end{align}
where $\gamma_m=4/\beta_0$, $\beta_0=11-(2/3)n_f$, 
$\tau={\rm e}^2-1$, and ${\cal F}(y) = \{1 - \exp(-y/[\Lambda_I^2])\}/y$, $\Lambda_I=1\,$GeV.  %
In the lighter quark sector, $(\mathpzc l,s)$ \cite{Qin:2011xq}:
$n_f=4$, $\Lambda_{\rm QCD} = 0.234\,$GeV, $\omega=0.5\,$GeV, $\omega D = (0.8\,{\rm GeV})^3$.
For the heavy quarks, $(c, b)$ \cite{Qin:2018dqp}: $n_f=5$, $\Lambda_{\rm QCD} = 0.36\,$GeV, $\omega=0.8\,$GeV, $\omega D = (0.6\,{\rm GeV})^3$.
In each sector, with the product $\omega D$ held fixed, predictions are practically independent of variations $\omega \to \omega (1\pm 0.1)$; so, the representation is solid.

Using Eq.\,\eqref{defcalG}, the kernel of the quark gap equation can be written as 
\cite{Maris:1997tm, Binosi:2016wcx}:
\begin{equation}
\label{DSEkernel}
{\mathpzc d}(l^2) T_{\mu\nu}^{l} [i\gamma_\mu\frac{\lambda^{a}}{2} ] \otimes [i\gamma_\nu\frac{\lambda^{a}}{2} ] \,,
\end{equation}
We use a mass-independent momentum-subtraction renormalisation scheme, with each renormalisation constant fixed in the chiral limit  \cite{Chang:2008ec}.
Using standard procedures \cite{Munczek:1994zz, Bender:1996bb}, the Bethe-Salpeter kernel is immediately obtained from Eq.\,\eqref{DSEkernel}.

It remains only to specify the renormalisation group invariant (RGI) quark current masses \cite{Politzer:1976tv}, which we denote as $\hat m_f$, $f=\ell, s, c, b$.
These are the primary expressions of the effect of Higgs boson couplings into QCD: they are independent of renormalisation scale and scheme; so, can be used to compare results across different theoretical frameworks and experiments.
We take the values determined elsewhere \cite{Yao:2021pyf, Yao:2021pdy}, all of which are listed in Table~\ref{RGImasses} and match well with those recorded in other sources.

\begin{table}[t]
\caption{\label{RGImasses}
Quark masses. 
Row 1.  Values of RGI current masses, from which all other listed results can be obtained.
Row 2. RGI masses evolved to $2\,$ GeV, the scale that is often quoted.
Row 3. Euclidean constituent quark mass \cite[Eq.\,(58)]{Maris:1997tm}, which is a nonperturbative extension of the ``running mass'' used in Ref.\,\cite[PDG]{ParticleDataGroup:2024cfk}.
%
}
\begin{center}
\begin{tabular*}
{\hsize}
{
l@{\extracolsep{0ptplus1fil}}
|c@{\extracolsep{0ptplus1fil}}
c@{\extracolsep{0ptplus1fil}}
c@{\extracolsep{0ptplus1fil}}
c@{\extracolsep{0ptplus1fil}}}\hline\hline
GeV            $\ $ & $\ell=u,d\ $ & $s$ & $c\ $ & $b\ $ \\  \hline
RGI $\ $ & $0.0068\ $ & $0.168\ $ & $1.61\ $ & $6.96\ $ \\
1-loop $2\,$GeV$\ $  & $0.0044\ $ & $0.112\ $ & $1.22\ $ & $5.26\ $ \\
Eucl.\ CQM$\ $ & $0.405\phantom{0}\ $ & $0.560\ $ & $1.32\ $ & $4.12\ $ \\
\hline\hline
\end{tabular*}
\end{center}
\end{table}

Physically, the weak transition vertex, ${\mathpzc W}_\mu^{cb}$, must exhibit poles at timelike total momenta.
For $\mathsf P \to \mathsf P_f$, only the vector part of this vertex contributes to semileptonic transitions.  
It has poles at $Q^2 + m_{B_c^\ast,B_{c0}^\ast}^2=0$.
In $\mathsf P \to \mathsf V_f$ transitions, the axial-vector part also contributes: 
it has poles at $Q^2 + m_{B_c,  B_{c1}}^2=0$.  
The expression of such poles is necessary for any valid analysis of the transitions discussed herein.  They are manifest in our treatment.

\smallskip

\noindent\emph{4.\ Methods}\,---\,%
%
The integration in Eq.\,\eqref{dMDs} samples the integrand functions on a significant domain of their complex-valued arguments.  
When the masses of the initial and final state mesons are similar, \emph{i.e}.\ the ratio of the current-masses of the quarks involved, $r_{Q_1:Q_2}$, does not much differ from unity, then $t_-$ remains modest.
In such cases, the integral can readily be evaluated using standard algorithms \cite{Maris:1997tm, Maris:2005tt, Krassnigg:2009gd}.  
However, at some value of $r_{Q_1:Q_2}=:r_f$, $t_-$ becomes so large that singularities associated with the analytic structure of the quark propagators \cite{Maris:1997tm} move into the complex-${\mathpzc q}^2$ integration domain; then straightforward numerical techniques fail.
To circumvent this issue, we adopt the method in Ref.\,\cite{Yao:2021pyf}.

\smallskip

\noindent (\emph{I}) \ldots 
We consider the semileptonic transitions of a fictitious $b \bar{\mathpzc Q}$ pseudoscalar meson: 
$\bar B_{b \bar {\mathpzc Q}} \to \mathsf P_{f b \bar {\mathpzc Q}}, \mathsf V_{f b \bar {\mathpzc Q}}$.
All relevant Schwinger functions (propagators and vertices) and, subsequently, the transition form factors, are calculated as a function of $\hat m_{\mathpzc Q}$ as it is decreased from $r_{{\mathpzc Q}:b}=1$ to $r_{{\mathpzc Q}:b}^f=0.47$, which is the minimum value achievable via the standard algorithms. 
Then, using the Schlessinger point method (SPM) -- see, \emph{e.g}.,  Refs.\,\cite{PhysRev.167.1411, Schlessinger:1966zz, Tripolt:2016cya, Yao:2024ixu, Cheng:2024cxk} -- we develop $\hat m_{\mathpzc Q}$-interpolations of all transition form factors, which are then employed to extrapolate every measurable feature of the matrix elements to the physical point $r_{u:b}=0.001$.
Of course, $r_{s:b}=0.024$ is reached along the way.
Naturally, in completing these calculations, one also obtains the masses of all mesons connected with the processes via their respective Bethe-Salpeter equations.

\smallskip

\noindent (\emph{II}) \ldots 
Next, these exercises are repeated from the reverse perspective.  
Namely, beginning with a fictitious light initial state, we consider 
$B_{{\mathpzc Q}\bar {\mathpzc l}}
\to \mathsf P_{f {\mathpzc Q}\bar{\mathpzc l}}, \mathsf V_{f {\mathpzc Q} \bar {\mathpzc l}}$ transitions.
The form factors are then calculated as a function of $\hat m_{\mathpzc Q}$, 
increasing it from the point $r_{\mathpzc l:{\mathpzc Q}}=1$ to that for which $r_f^{\bar {\mathpzc l}:{\mathpzc Q}}=0.007$, \emph{i.e}., the smallest achievable value on this trajectory.
The same is done for $B_{{\mathpzc Q}\bar s}
\to \mathsf P_{f {\mathpzc Q}\bar s}, \mathsf V_{f {\mathpzc Q} \bar s}$ transitions, for which one can reach $r_f^{\bar s:{\mathpzc Q}}=0.15$.
Then, SPM extrapolation is used to reach the physical values 
$r_{\bar {\mathpzc l}:b}=0.001$, $r_{\bar s:b}=0.024$ 
for all transition form factors. 
Meson masses are also obtained via this route.

\smallskip

\noindent (\emph{III}) \ldots Upon completion of these exercises, to produce final results, we combine the outcomes in the manner detailed below.

\smallskip

The SPM employs interpolation via continued fractions \cite{Schlessinger:1966zz, PhysRev.167.1411}.  
It is typically augmented by statistical sampling.   
The method avoids any assumptions about the function used for the representation of source information and incorporates and expresses both local and global features of that source.  
The latter property underpins the reliability of subsequent extrapolations.  
The SPM accurately reconstitutes a complex-valued function within a radius of convergence determined by that one of the function's branch points which lies nearest the real domain from which the sample points are drawn.  
The statistical aspect guarantees that one has a sound estimate of the uncertainty associated with any extrapolation.

In implementing the SPM, we first calculate the value of a given quantity, ${\mathpzc X}$, at $N=40$ different values of the changing mass, $\hat m_{\mathpzc Q}$, distributed evenly on the direct computation domain.
Then, $M=20$ values of $\hat m_{\mathpzc Q}$ are chosen at random from this $40$-element set, wherewith a continued fraction interpolation of ${\mathpzc X}(\hat m_{\mathpzc Q})$ is built on this $20$-element subset.
A large number of interpolating functions, $n_I$, is subsequently obtained by reviewing the $C(N,M)$ combinatorial possibilities for the $M$ element subset and removing those functions that fail to satisfy a simple physical constraint, \emph{viz}.\ we insist that each interpolation be smooth on the entire domain of required current-quark masses.  
For all relevant quantities, this constraint yields $n_I\approx 100,000$ acceptable interpolations.
The result for ${\mathpzc X}$ is then obtained by extrapolating each of the $n_I$ accepted SPM interpolants to the target current-mass and reporting the value that sits at the centre of the band within which 68\% of the interpolants lie.  
The associated $1\sigma$ band is quoted as the uncertainty.

\begin{table}[t]
\caption{\label{mesonmasses}
Predicted masses (in GeV) of the ten mesons that play a role in the transition form factors discussed herein. 
PDG does not list empirical values for masses of $B_c^\ast$, $B_{c0}^\ast$, $B_{c1}$.
Nevertheless, Ref.\,\cite[ATLAS]{ATLAS:2026ubk} reports $m_{B_c^\ast} = 6.339(187)\,$GeV.
%
}
\begin{center}
\begin{tabular*}
{\hsize}
{
l@{\extracolsep{0ptplus1fil}}
c@{\extracolsep{0ptplus1fil}}
| c@{\extracolsep{0ptplus1fil}}
c@{\extracolsep{0ptplus1fil}}}\hline\hline
meson$\ $ & $J^P\ $ & herein $\ $ & \cite[PDG]{ParticleDataGroup:2024cfk} \\ \hline
$B$          & $0^-$ & 5.27(32)  & $5.28\ $ \\ 
$D$          & $0^-$ & 1.87(05)   & $1.87\ $ \\ 
$B_s$        & $0^-$ & 5.39(35)  & $5.37\ $ \\ 
$D_s$        & $0^-$ & 1.97(06)   & $1.97\ $ \\ 
$D^\ast$        & $1^-$ & 2.03(09)   & $2.01\ $ \\ 
$D_s^\ast$      & $1^-$ & 2.16(08)   & $2.11\ $ \\ 
$B_c$        & $0^-$ & 6.27(26)  & $6.27\ $ \\ 
$B_c^\ast$      & $1^-$ & 6.37(03) & \\ 
$B_{c0}^\ast$   & $0^+$ & 6.75(03) & \\ 
$B_{c1}$     & $1^+$ & 6.87(03) & \\ 
\hline\hline
\end{tabular*}
\end{center}
\end{table}

The reliability of this procedure can be illustrated via results for the meson masses, which are listed in Table~\ref{mesonmasses}.  
Both (\emph{I}) and (\emph{II}) yield values which all agree within mutual uncertainties.
Furthermore, the mean relative difference between the values, as measured against each average, is $0.85(63)$\%.
So, they can be combined to deliver central results and associated uncertainties.

Regarding the transition form factors, we note that on the physical domain associated with any value of $r_{Q_1:Q_2}$, each can  accurately be interpolated using  \cite{Yao:2021pyf}:
\begin{equation}
\label{TFfunction}
{\mathpzc f}(t) = \alpha_1 +  \alpha_2 \, t+ \frac{\alpha_3\,  t^2}{{\mathpzc m}^2-t}\,,
\end{equation}
where $\alpha_{1,2,3}$ and ${\mathpzc m}$ are functions of $r_{Q_1:Q_2}$.  
Referring to Eq.\,\eqref{EqMEs}, the pole masses in Eq.\,\eqref{TFfunction} correspond to particular mesons.
For ${\mathpzc f}=A_0$, ${\mathpzc m}=m_{B_c}$;
${\mathpzc f}=f_+, V$, then ${\mathpzc m}=m_{B_c^\ast}$;
${\mathpzc f}=f_0$, then ${\mathpzc m}=m_{B_{c0}^\ast}$;
and
${\mathpzc f}=A_{1,2}$, then ${\mathpzc m}=m_{B_{c1}}$.
These masses have already been calculated; see Table~\ref{mesonmasses}.   
So, it is only necessary to develop SPM interpolations of the coefficients $\alpha_{1,2,3}$.
Once again, in all cases, both (\emph{I}) and (\emph{II}) yield mutually compatible results, which can therefore be averaged; see  Table~\ref{SPMparameters}.

\begin{table}[t]
\caption{\label{SPMparameters}
Referring to Eq.\,\eqref{TFfunction}, SPM interpolation parameters for each transition form factor considered herein.
}
\begin{center}
\begin{tabular*}
{\hsize}
{
l@{\extracolsep{0ptplus1fil}}
c@{\extracolsep{0ptplus1fil}}
|c@{\extracolsep{0ptplus1fil}}
c@{\extracolsep{0ptplus1fil}}
c@{\extracolsep{0ptplus1fil}}}\hline\hline
transition$\ $ & $\mathpzc f\ $  & $\alpha_1^{\mathpzc f}\ $ & $\alpha_2^{\mathpzc f}\times{\rm GeV}^2\ $ & $\alpha_3^{\mathpzc f}\times {\rm GeV}^2\ $ \\ \hline
$B\to D$       & $f_+\ $ & 0.644(15) & 0.0316(12) & 0.0441(30) \\
               & $f_0\ $ & 0.644(15) & 0.0122(11) & 0.0145(24) \\\hline
$B_s\to D_s$   & $f_+\ $ & 0.672(16) & 0.0317(15) & 0.0463(28) \\
               & $f_0\ $ & 0.672(17) & 0.0106(14) & 0.0184(24) \\ \hline
$B\to D^\ast$     & $A_0\ $ & 0.560(19) & 0.0347(18) & 0.0561(11) \\
               & $A_1\ $ & 0.554(13) & 0.0114(15) & 0.0405(25) \\
               & $A_2\ $ & 0.542(19) & 0.0307(27) & 0.0751(27) \\
               & $V\ $   & 0.698(21) & 0.0268(17) & 0.0862(15) \\ \hline
$B_s\to D_s^*$ & $A_0\ $ & 0.632(16) & 0.0263(18) & 0.0433(12) \\
               & $A_1\ $ & 0.597(13) & 0.0141(15) & 0.0262(10) \\
               & $A_2\ $ & 0.552(27) & 0.0213(14) & 0.0505(16) \\
               & $V\ $   & 0.740(12) & 0.0416(14) & 0.0657(12) \\
\hline\hline
\end{tabular*}
\end{center}
\end{table}

\begin{figure*}[t]
\begin{center}
\begin{tabular}{lcl}
\includegraphics[clip,width=0.45\linewidth]{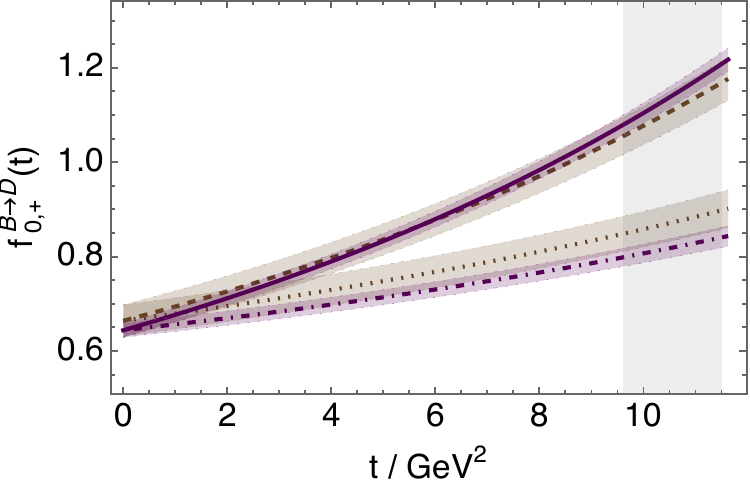}\hspace*{2ex } & \hspace*{2em}  & 
\includegraphics[clip,width=0.45\linewidth]{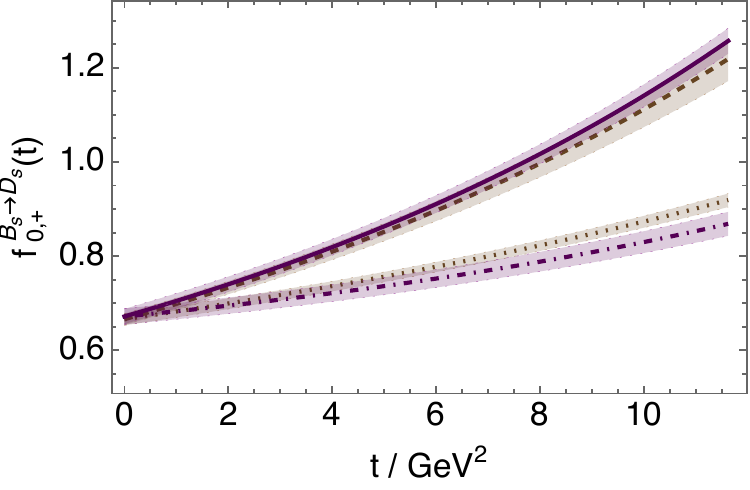}\vspace*{-40ex} \\
{\large\sf A} & &  {\large\sf B} 
\end{tabular}
\vspace*{34ex}
\end{center}

\begin{center}
\begin{tabular}{lcl}
\includegraphics[clip,width=0.45\linewidth]{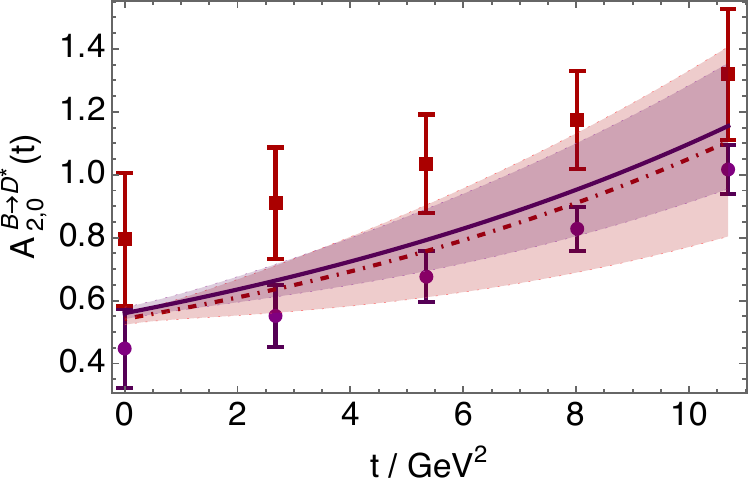}\hspace*{2ex } & \hspace*{2em}  & 
\includegraphics[clip,width=0.45\linewidth]{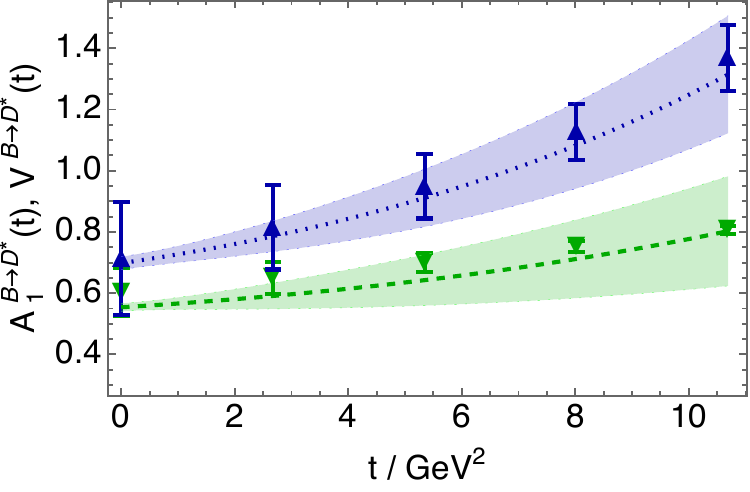}\vspace*{-40ex} \\
{\large\sf C} & &  {\large\sf D} 
\end{tabular}
\vspace*{34ex}
\end{center}

\begin{center}
\begin{tabular}{lcl}
\includegraphics[clip,width=0.45\linewidth]{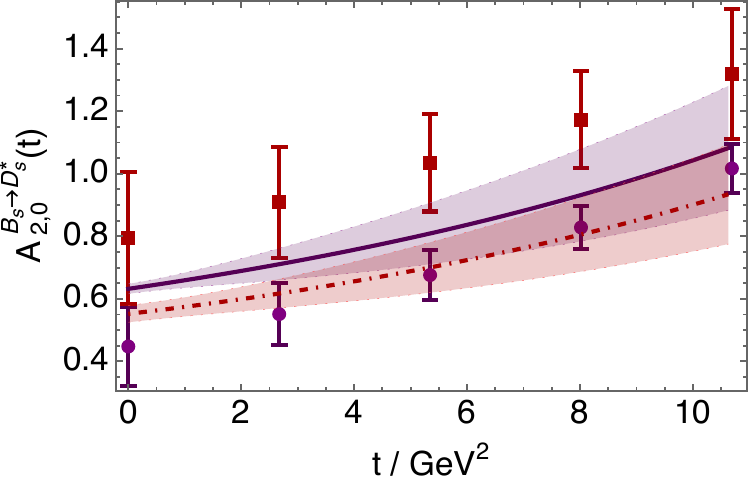}\hspace*{2ex } & \hspace*{2em}  & 
\includegraphics[clip,width=0.45\linewidth]{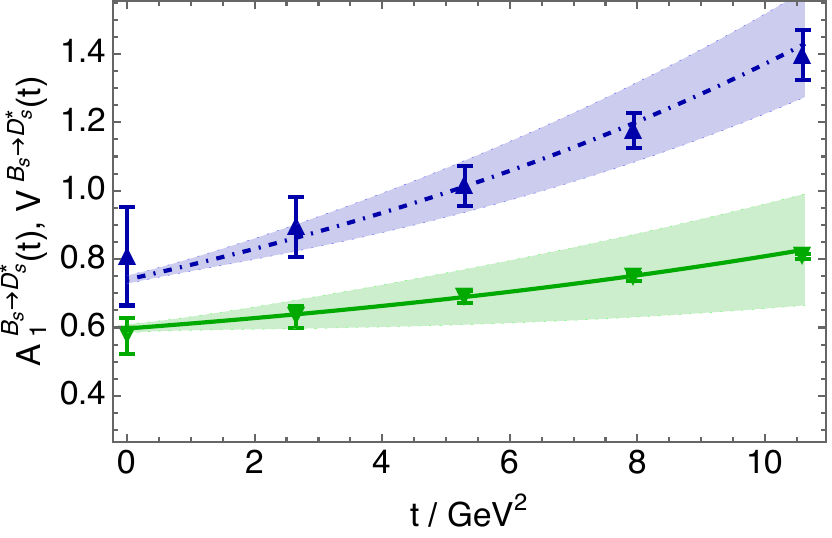}\vspace*{-40ex} \\
{\large\sf E} & &  {\large\sf F} 
\end{tabular}
\vspace*{32ex}
\end{center}

\caption{\label{F2}
{\sf Panel A}.\ 
$B\to D$ semileptonic transition form factors: 
$f_+$ -- solid purple curve; $f_0$ -- dot-dashed purple curve.
In both cases, the associated band indicates the $1\sigma$ SPM uncertainty.
Comparisons.\   
$f_+$ -- dashed brown curve; $f_0$ -- dotted brown curve.
Extrapolations using fits to four points (evenly spaced in $t$ within vertical grey band) computed via numerical simulation of lattice-regularised QCD (lQCD) with the $b$ quark treated as nonrelativistic \cite{Na:2015kha}.
{\sf Panel B}.\ 
$B_s\to D_s$ transition form factors: 
$f_+$ -- solid purple curve; $f_0$ -- dot-dashed purple curve.
Again, the associated bands indicate the $1\sigma$ SPM uncertainties.
Comparisons.\   
$f_+$ -- dashed brown curve; $f_0$ -- dotted brown curve.
Extrapolations using fits to three/four lQCD points computed at an array of unphysically light $b$ quark masses and subsequently extrapolated to the physical value \cite{McLean:2019qcx}.  For $f_+$, the three points are eventually evenly spaced on $0 \lesssim t/{\rm GeV}^2 \lesssim 5$; and for $f_0$, the coverage of the four evenly spaced points finally grows to $0 \lesssim t/{\rm GeV}^2 \lesssim 9$.
{\sf Panels C and D}.\ $B \to D^\ast$ transition form factors. 
Legend. 
Curves: $A_2$ -- solid purple; $A_0$ -- dot-dashed red; $A_1$ -- dashed green; $V$ -- dotted blue.  Like coloured bands are the $1\sigma$ SPM uncertainty.
Points. Taken from Ref.\,\cite{Harrison:2023dzh} and representative of the lQCD analysis described therein, where one can also find details about the lattice setup and form factor inference methods.
{\sf Panels E and F}.\ $B_s \to D_s^\ast$ form factors.   Legend as in {\sf Panels C and D}.
}
\end{figure*}

\smallskip

\noindent\emph{5.\ Form Factor Results}\,---\,%
%
Our predictions for the $B_{(s)} \to D_{(s)}$ semileptonic transition form factors are depicted in Figs.\,\ref{F2}A, \ref{F2}B.
Forming the ratios of kindred functions, one obtains curves that are practically $t$-independent.
For $f_+$, the $B_s\to D_s/B\to D$ ratio is $1.036(1)$; and for $f_0$, it is $1.032(3)$.  
Evidently, there is a uniform $\approx 3$\% violation of SU$(3)$-flavour symmetry.

Working with the plotted form factors, one obtains the partial widths and branching fractions reported in Table~\ref{BranchingFracs}: our predictions are in excellent agreement with all available empirical results.  
Considering the light-fermion final states, the signal for violation of SU$(3)$-flavour symmetry is $9(12)$\%.

The CSM predictions for $B_{(s)} \to D_{(s)}^\ast$ semileptonic transition form factors are depicted in Figs.\,\ref{F2}C\,-\,\ref{F2}F.
An indication of the level of SU$(3)$-flavour symmetry violation can immediately be read from the $t=0$ $D_s^\ast/D^\ast$ ratios of form factor values: 
$A_0$ -- $1.13(4)$; 
$A_1$ -- $1.08(3)$; 
$A_2$ -- $1.02(6)$; 
$V$ -- $1.06(3)$.

Before providing predictions for the partial widths and branching fractions, it is worth discussing the comparisons made in Figs.\,\ref{F2}C\,-\,\ref{F2}F between results obtained using continuum and lattice Schwinger function methods.  
(The lQCD study obtained form factors at six independent heavy quark masses below that of the $b$ quark and extrapolated to the physical point.  Other lattice setup and algorithm details are provided in Ref.\,\cite{Harrison:2023dzh}.)
%
%
Regarding Figs.\,\ref{F2}D, \ref{F2}F, \textit{i.e}., $A_1$, $V$, CSM and lQCD are in agreement.  
This is not the case for $A_{0,2}$ -- see Figs.\,\ref{F2}C, \ref{F2}E: CSM and lQCD results even disagree on the ordering of magnitudes for these form factors.  
The issue may lie with the fact that the lQCD extraction of $A_{0,2}$ is very uncertain.  
Indeed, as stressed in connection with Ref.\,\cite[Appendix~E, Fig.\,27]{Harrison:2023dzh},
the answers one obtains from lQCD are very sensitive to the method used in proceeding from basic lQCD outputs to the results interpreted as physical form factor values.  (\emph{N.B}.\ Our $A_{0,2}$ map closely into the lQCD-reported $h_{A_2,A_3}$.) 
Our ordering of $A_{0,2}$ has no such sensitivity and it agrees, too, with that predicted by the entirely algebraic analysis described in Ref.\,\cite{Xing:2022sor}.

Working with the form factors in Figs.\,\ref{F2}C\,-\,\ref{F2}F, one obtains the partial widths and branching fractions in Table~\ref{BranchingFracs}.
Considering the light-fermion final states, the signal for violation of SU$(3)$-flavour symmetry is $20(15)$\%, which is twice as large as that in the case of pseudoscalar meson final states.
Notably, in nearly all cases where comparison with empirical results is possible, our predictions are in agreement, \emph{viz}.\ within $1.5\sigma$.
The standout discrepancy is $B \to D^\ast \tau \bar\nu_\tau$: our prediction and the listed PDG result disagree at the level of $3\sigma$.

\smallskip

\noindent\emph{6.\ Lepton Flavour Universality}\,---\,%
%
We can now deliver predictions for the LFU ratios defined in Eq.\,\eqref{BFratios}:
\begin{subequations}
\begin{align}
\{ R(D) \, , \, R(D)^\ast \} & = \{ 0.290(11) \, , \, 0.267(5) \} \,,  \label{RDDs}
\\
\{ R(D_{s}) \, , \, R(D_{s}^{\ast}) \}  & = \{ 0.283(11) \, , \, 0.251(5) \} \,. \label{RDsDss}
\end{align}
\end{subequations}

The result in Eq.\,\eqref{RDDs} is plotted in Fig.\,\ref{F1F}: our parameter free prediction agrees with the most recent available estimate produced by the ``flavor lattice averaging group'' \cite[FLAG]{FlavourLatticeAveragingGroupFLAG:2024oxs}.
Regarding the Ref.\,\cite[HFLAV]{HFLAV:2024ctg} estimate, our prediction differs by $1.3\sigma$; so, it is compatible.

\begin{figure}[t]
\centerline{%
\includegraphics[clip, width=0.44\textwidth]{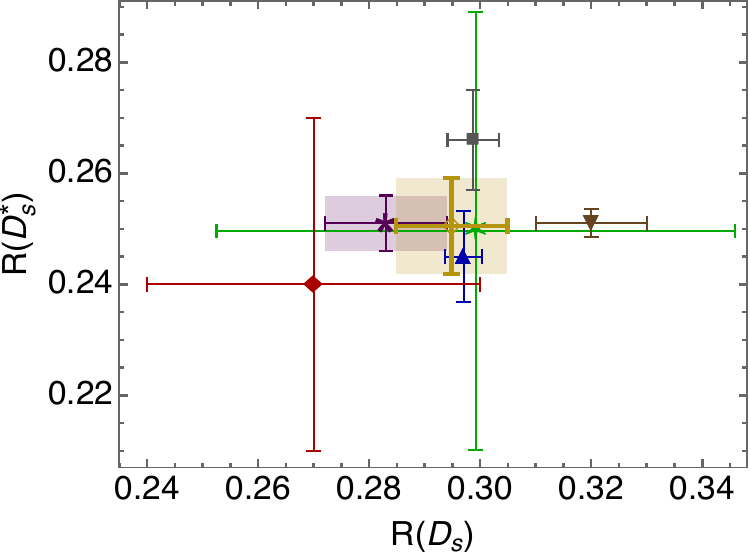}}

\vspace*{-2ex}

\caption{\label{F3}
CSM prediction for $\{R(D_s) , R(D_s^\ast)\}$ -- purple star within uncertainty-bounding rectangle.
Comparisons with selected independent, data-unbiased estimates.
Legend.
Green cross -- light-front quark model \cite[CLFQM]{Kang:2018jzg}.
Brown down triangle -- instantaneous Bethe-Salpeter equation model \cite[IBSE]{Zhou:2019stx}.
Blue up triangle -- heavy quark expansion phenomenology \cite[HQE]{Bordone:2019guc}.
Grey square -- simulation of lQCD \cite[lQCDGl]{Harrison:2023dzh}.
Red diamond -- covariant confining quark model \cite{Ivanov:2026pul}.
Gold open diamond -- unweighted average of plotted results.
}
\end{figure}

No such comparisons are available for the prediction in Eq.\,\eqref{RDsDss}.  
On the other hand, as noted at the outset, independent, data-unbiased estimates are available from an array of theory tools \cite{Kang:2018jzg, Zhou:2019stx, Bordone:2019guc, Harrison:2023dzh, Ivanov:2026pul}.  
They are plotted in Fig.\,\ref{F3}.  
Their unweighted average is 
$\{ R(D_{s}) \, , \, R(D_{s}^{\ast}) \}  = \{ 0.297(12) \, , \, 0.250(10) \}$.

Our prediction, Eq.\,\eqref{RDsDss}, is also drawn in Fig.\,\ref{F3}.  
Including it in the average, one obtains the value reported in Eq.\,\eqref{ResRDsDss}:
the parameter-free CSM result is practically definitive of the mean of available estimates.
Compared with our prediction for $\{ R(D) \, , \, R(D)^\ast \} $, we see a $2.3\sigma$ signal for SU$(3)$-flavour symmetry violation.

\smallskip

\noindent\textit{7.\ Perspective and Summary}\,---\,%
The search for physics beyond the Standard Model (SM) is increasingly being pursued at the intensity frontier, with accelerators worldwide attempting to gather high-quality data that can be used to test SM predictions with unprecedented precision.
Such indirect searches for new phenomena can potentially serve as sensitive testing grounds in the absence of new, more powerful colliders.

The analysis and predictions presented herein impact upon the SM flavour sector, which is fruitful ground for SM verification; in particular, in the area of lepton flavour universality (LFU).   
Focusing on semileptonic heavy-to-heavy meson decays, which see $b$ quarks transform into $c$ quarks, we did not attempt to constrain the associated Cabibbo-Kobayashi-Maskawa (CKM) matrix element, $|V_{cb}|$.  
Instead we focused on quantities that are independent of $|V_{cb}|$ and used continuum Schwinger function methods to deliver entirely parameter free results for the twelve strong-interaction form factors that are key to evaluating four measurable LFU ratios.
Therewith, we find a $2.8\sigma$ SM theory vs.\ experiment tension in the light-quark sector.  
This is less than the typically quoted mismatch because our prediction for the $D^\ast$ ratio, $R(D^\ast)$, is larger than that commonly used to measure the tension.

Furthermore, our predictions for the strange-quark-spectator LFU ratios straddle the averages of an array of disparate theory estimates and possess somewhat smaller uncertainties.  
They also indicate 5\% SU$(3)$-flavour symmetry violation in these transitions with $2.3\sigma$ (99\%) confidence.  
This is the same size as found in the the $D^\ast_s-D^\ast$ mass splitting and is typical of that produced by Higgs boson couplings into the SM strong interaction sector after modulation by emergent phenomena driven by strong interaction gauge-boson (gluon) dynamics.

\smallskip

\noindent\textit{Acknowledgments}\,---\,%
%
Work supported by
National Natural Science Foundation of China, grant no.\ 12135007.

\smallskip



\end{document}